\documentclass[prc,twocolumn,pdflatex,nofootinbib,superscriptaddress,floatfix]{revtex4-1}
\usepackage[T1,T2A]{fontenc}
\usepackage[utf8]{inputenc}
\usepackage[english]{babel}
\usepackage{graphicx}
\usepackage{bm}
\usepackage{placeins}
\usepackage{xspace}
\usepackage{bm,amsmath,amssymb,amsfonts}
\usepackage{comment}
\usepackage{hyperref}
\hypersetup{
    colorlinks=true,
    linkcolor=blue,
    citecolor=blue,
    filecolor=blue,      
    urlcolor=blue,
}
\usepackage{bm,amsmath,amssymb,amsfonts}
\newcommand{\eq}[1]{\begin{align} #1 \end{align}}

\newcommand{\be}{\begin{equation}}
\newcommand{\ee}{\end{equation}}

\usepackage{color}

\graphicspath{ {/tmp/} }

\begin{document}

\title{Thermodynamic properties of interacting bosons \\ 
with zero chemical potential}

\author{O. S. Stashko}
\affiliation{ Taras Shevchenko National University of Kyiv, 03022 Kyiv, Ukraine}
\author{D. V. Anchishkin}
\affiliation{
Taras Shevchenko National University of Kyiv, 03022 Kyiv, Ukraine}
\affiliation{
Bogolyubov Institute for Theoretical Physics, 03680 Kyiv, Ukraine}
\author{O. V. Savchuk}
\affiliation{
Taras Shevchenko National University of Kyiv, 03022 Kyiv, Ukraine}
\author{M. I. Gorenstein}
\affiliation{
Bogolyubov Institute for Theoretical Physics, 03680 Kyiv, Ukraine}

\begin{abstract}
\normalsize
Thermodynamics properties of an interacting system of bosons are considered at finite temperatures and zero chemical potential within the Skyrme-like mean-field model. An interplay between attractive and repulsive interactions is investigated.
As a particular example an equilibrium system of  pions is discussed. 
Several modifications of  thermodynamic properties in the considered system are found with  increasing a strength of attractive forces.  Different types  of the first order phase transition are classified. Some of these transitions exist also  
in the Boltzmann approximation. 
However, effects of the Bose statistics introduce the notable additional changes in the thermodynamic quantities due to a possibility of the Bose-Einstein condensation.
\end{abstract}
\maketitle
\section{Introduction}
The phase structure of the strongly interacting matter is currently a subject of active researches \cite{busza2018heavy, Bzdak_2020}. At low temperatures and densities the system  behaves as the ideal gas mixture of different particle species. The equation of state (EoS) at higher temperature and zero baryonic density is of particular interest as it can be studied within the lattice QCD numerical methods \cite{PhysRevD.80.014504,Bors_nyi_2010}. A special attention was also devoted to the mesonic degrees of freedom~\cite{PhysRevLett.86.592, BIRSE200127, Mannarelli_2019,RevModPhys.50.107}. At small temperature and zero chemical potentials a main contribution to the EoS is expected from pions which are the lightest particles with strong interactions. Pions are bosons, thus, some specific effects connected to the Bose statistics are also possible. 

Effects of interactions for the thermodynamic properties of many-body boson system were discussed in some detail (see, e.g., Ref.~\cite{kapusta_gale_2006}). A characteristic trait of such systems is a phenomenon of the Bose-Einstein condensation (BEC)~\cite{Bose:1924mk,einstein1925stizunger}. 
The BEC of pions in high energy collisions were discussed in Refs.~\cite{Begun:2006gj,Begun:2008hq}.
The BEC can  be also expected 
for other types of  bosons with the strong interactions, e.g., kaons and $\alpha$ nuclei, and applied to study properties of nuclear matter,  boson stars, etc~\cite{Liebling2012,Brown_1994,Li_1997, Brandt:2018bwq, andersen2018boseeinstein, Umeda1994,Satarov:2017jtu,Strinati_2018,Nozieres:1985zz,PhysRevLett.101.082502,Chavanis:2011cz,
Mishustin_2019}. 
Recent works investigate a possibility of the BEC for pion with non-zero isospin  using  the lattice QCD formulation~\cite{Brandt:2017oyy,Brandt_2018}. It was also studied  in the phenomenological and effective theories in Refs.~\cite{savchuk2020boseeinstein,Anchishkin_2019,Kogut_2001,Toublan_2003,Mammarella_2015,Carignano_2017}.

Particle interactions, when both attractive and repulsive effects are included,
can lead to the first order liquid-gas phase transitions (LGPT) that is a
common features of all molecular systems. The possibility of first-order phase transition via formation of the BEC was introduced in Ref.~\cite{Anchishkin_2019}.

In the present paper we follow this idea by studying an interplay  between two physical phenomena in the system of interacting bosons: LGPT and BEC. The system of bosons with zero chemical potential is considered, and the mean-field model  with the Skyrme-like interaction is adopted. This form of the mean-field interactions was discussed earlier in Refs.~\cite{Satarov2017,Anchishkin_2019}.
Different types of the LGPT in an interacting boson system will be considered.
By increasing a strength of an attractive part of the mean-field potential
we find qualitatively different scenarios: no LGPT and no BEC, LGPT without
BEC, LGPT with BEC, and unstable vacuum.
Our analysis is given in a general form which is valid for any system of bosons,
whereas our numerical calculations are referred to interacting pions as
a generic example.

The paper is organized as follows. Section~\ref{sec:MeanField} gives a general description of the mean-field framework with the Skyrme-like interaction. In  Sec.~\ref{sec:EoS} possible scenarios for the system EoS are presented. 
In Sec.~\ref{sec:fluc} the particle number fluctuations are calculated in different thermodynamic scenarios.
Section~\ref{sec:sum} presents a short summary.

\section{mean field Skyrme model}\label{sec:MeanField}
The thermodynamic mean-field framework for a system of interacting bosons will be defined as a set of the following  coupled equations for the pressure $p$ and particle number density $n$ (see, e.g., Ref.~\cite{Anchishkin:2014hfa}): 
\eq{\label{MF-p}
p(T,\mu)&=p_{\rm id}(T,\mu^{*})+\int\limits_{0}^{n}dn'\,n' \frac{dU(n')}{dn'}~,\\
\label{MF-n}
n&=\left(\frac{\partial p}{\partial \mu}\right)_T = n_{\rm id}(T,\mu^{*})~,\\
\label{MF-mu}
\mu^* & = \mu-U(n)~,
}
where $T$ and $\mu$ are the system temperature and chemical potential, respectively.
A density dependent mean-field potential $U(n)$ will be taken in the Skyrme-like form as
(see, e.g., Ref.~\cite{Satarov:2017jtu,Anchishkin_2019}):
\eq{\label{Un}
U(n)~=~-An+Bn^2~.
}
Positive constants $A$ and $B$ describe, respectively,  attractive and repulsive effects of particle interactions.

In Eqs.~(\ref{MF-p})-(\ref{MF-n}),  $p_{\rm id}$ and $n_{\rm id}$ are the pressure and particle number density of the ideal gas in the grand canonical ensemble \cite{GNS}
\eq{\label{p-id}
p_{\rm id}(T,\mu^*)&=\frac{ g}{6\pi^2} \int\limits_0^{\infty}
dk
\frac{ k^4}{\sqrt{k^{2} + m^2}}\, f_{\rm k}(T,\mu^*)~,\\
n_{\rm id}(T, \mu^{*}) & = \frac{g}{2\pi^2}\int\limits_{0}^{\infty}dk~k^2~f_{\rm k}(T,\mu^*)~,
\label{nid}
}
where $m$ is the particle mass,
and $g$ is the degeneracy factor.  The momentum distribution $f_k$ reads
\eq{\label{fk}
f_{\rm k}
(T,\mu^*)
=\left[{\rm \exp}\left(\frac{
\sqrt{k^2+m^2}-\mu^*}{T}\right)-\eta\,\right]^{-1}~,
}
where $\eta=1$ corresponds to the Bose statistics that will be discussed in our paper. The  $\eta=-1$ corresponds to the Fermi statistics, and $\eta=0$ to the Boltzmann approximation in which effects of a quantum statistics are neglected. 

For Bose particles the effective chemical potential $\mu^*$ (\ref{MF-mu}) is restricted from above by a particle mass, $\mu^*\le m$. 
At $\mu^*=m$ the BEC phenomenon takes place. In what follows the system of bosons with zero chemical potential $\mu=0$ is considered.  Thus, a condition of BEC reads
\eq{\label{cond-crit}
-~U(n)~=~m~.
}
Whether the condition (\ref{cond-crit}) can be satisfied  depends on the values of $A$ and $B$ parameters in Eq.~(\ref{Un}). Note that  $U(n)$ function (\ref{Un}) is a parabola with its lowest
negative value $-A^2/(4B)$ at $n=n_0=A/(2B)$. The condition (\ref{cond-crit}) can be therefore satisfied if $-U(n_0)\geq m$ that requires 
\begin{equation} \label{Acr}
 A \geq A_{\rm cr}~=~2\, \sqrt{B\,m}~.
\end{equation}
Thus, the BEC can be only possible if the attractive part of particle interactions is  strong enough. 
In this case,  Eq.~(\ref{cond-crit}) has two solutions, $n=n_1$ and $n=n_2$,
\eq{\label{n12}
n_{1,2}~=~\frac{A\pm\sqrt{A^2-4Bm}}{2B}~.
}
The BEC phenomenon can only occur at $n=n_1$ or at $n=n_2$.
It is not possible at $n<n_1$ and $n>n_2$. Besides, the particle densities $n$ inside the interval
$(n_1,n_2)$ are forbidden at any temperature $T$ as they lead to $\mu^*>m$ and negative values of the Bose-Einstein distribution function (\ref{fk})  at small momenta $k$. These properties of the system of bosons with the Skyrme-like $U(n)$ potential were recently noted in  Ref.~\cite{Anchishkin_2019}. 

\begin{figure}
    \centering
    \includegraphics[width=0.49\textwidth]{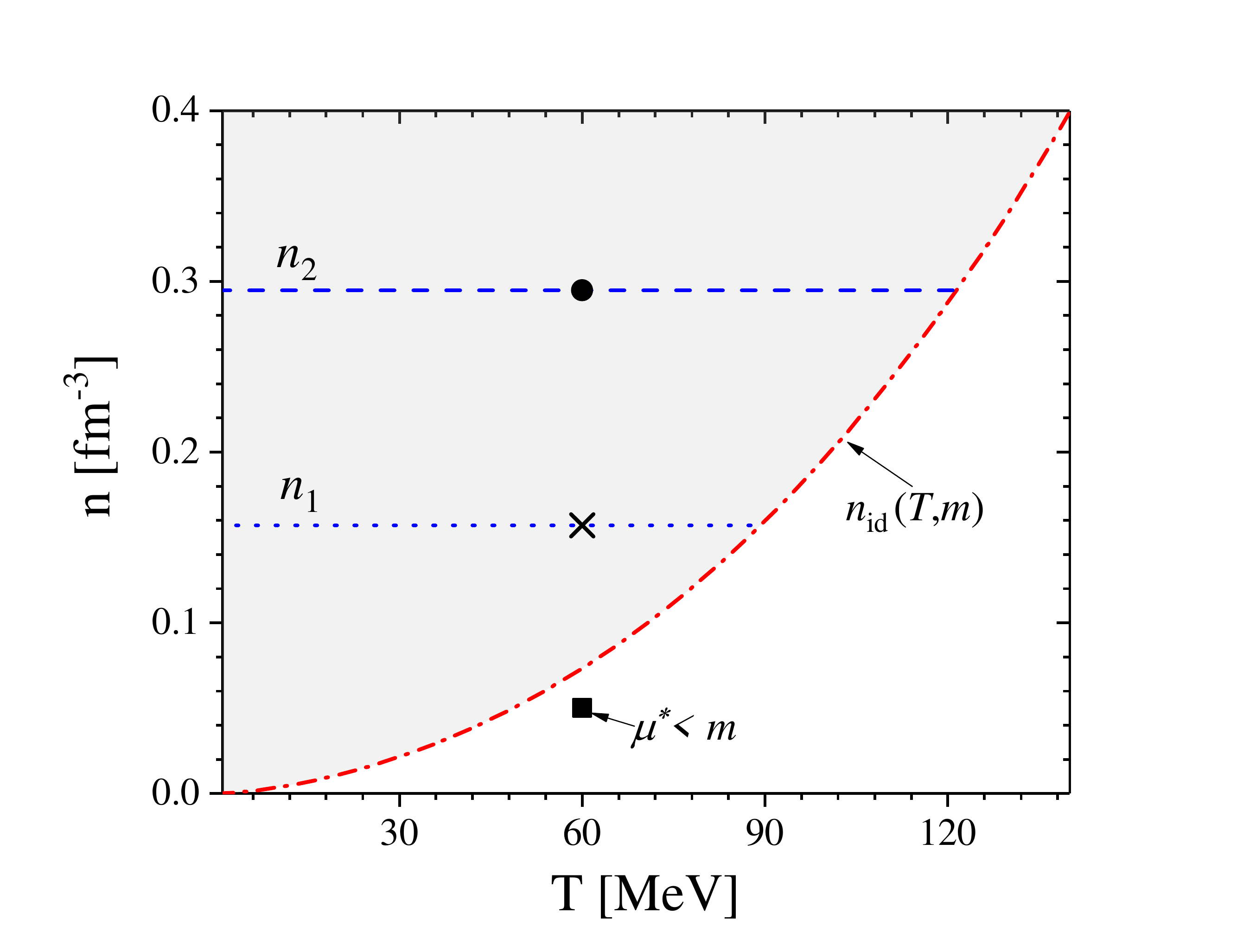}
    \caption{Possible states $n=n(T)$ at small $T$ at $A>A_{\rm cr}$. Dashed-dotted line shows the function $n=n_{\rm id}(T,\mu^*=m)$. Dotted and dashed horizontal lines present the states with $n=n_1$ and $n=n_2$, respectively. A square corresponds to $n=n_{\rm id}(T,\mu^*)$ with $\mu^*<m$. A circle and a cross present possible alternative states at the same $T$ with $\mu^*=m$ and $n_{\rm bc}>0$. The state with $n=n_1$ can only appear as an unstable one.  }
    \label{fig:BC}
\end{figure}

If $n(T)$ equal to either $n_1$ or $n_2$, Eq.~(\ref{MF-n}) should be extended as
\eq{\label{nbc}
n(T)~=~n_{1,2}~=~n_{\rm id}(T,\mu^*=m)~+~n_{\rm bc}~,
}
where $n_{\rm bc}\ge 0$ is the density of the Bose condensate (BC), i.e., a macroscopic part of  particles in the system of bosons that occupies a zero momentum level $k=0$. 
In Fig.~\ref{fig:BC} the $(T,n)$-plane is shown schematically in a case of $A>A_{\rm cr}$.
A dashed-dotted line in this figures presents the function $n_{\rm id}(T,\mu^*=m)$
The $(T,n)$ points under this line correspond to $\mu^*<m$, and no BC can be formed in this region. %
Thus, the function $n_{\rm id}(T,\mu^*=m)$ presents an upper limit of the particle number density at any $T$, if $n_{\rm bc}=0$.
Above the line $n_{\rm id}(T,\mu^*=m)$ the only admittable states are those given by Eq.~(\ref{nbc}) with $n_{\rm bc}>0$.    

However, a region of the particle number density $(n_1,n_2)$
is forbidden only for the \textit{pure} phases of the system. As will be discussed in the next sections, this region can be filled in by the liquid-gas \textit{mixed} phase with the `gas' density $n_g<n_1$ and `liquid' density $n_l=n_2$. At small $T$ we obtain the solutions  (\ref{MF-n}) with $\mu^*<m$. One of this solution is shown in Fig.~\ref{fig:BC} by the black square. For each of these solutions one should consider two alternative solutions $n(T)=n_1$ and $n(T)=n_2$ given by Eq.~(\ref{nbc}) and shown in Fig.~\ref{fig:BC} by the cross and circle, respectively. According to the Gibbs criterion one should choose a solution with the largest pressure. This solution is a stable one. Two other solutions correspond to metastable and unstable states. 
Note that the solution with $n(T)=n_1$ can only appear as an unstable one.
Indeed, by taking a difference of the pressures (\ref{MF-p}) at $\mu^*=m$ with and $n=n_2$ and  $n=n_1$ one finds
\eq{p(T,n_2)- p(T,n_1)=-\int\limits_{n_1}^{n_2}dn\,\left[ m+U(n)\right]~ >~  0~,
}
i.e., the pressure $p(T,n_2)$ is always larger than the pressure $p(T,n_1)$ at any $T$. 

\section{Equation of state and phase transitions}\label{sec:EoS}

In a Bose gas with the mean-field $U(n)$ discussed in the previous section  one obtains several  qualitatively distinct cases depending on the numerical values of the $A$ and $B$ parameters.
To be specific  these possibilities will be illustrated by choosing  a system of an interacting pions. We neglect  effects connected to electric interactions and small difference between the masses of neutral and charged pions. Therefore,  in what follows the particle  mass and degeneracy factor will be fixed as $m=m_\pi=140$~MeV and $g=3$.

To be even more specific, a fixed value of the $B$ parameter will be further considered.
At fixed chemical potential $\mu$ repulsion effects manifest themselves  in a suppression of the pressure, whereas attractive ones lead to an increase of the pressure. These slightly anti-intuitive results appear to be a general feature of a self consistent mean-field approach. A decrease or increase of the systems pressure is explained at fixed $\mu$ by the corresponding decrease or increase of particle number density due to, respectively, repulsive and attractive interactions. 
The $B$ parameter will be fixed as $B/m=21.6$~fm$^{-6}$. At $A=0$ this $B$ value leads to
a suppression of the pressure as
$p_{\rm id}(T,\mu^*)/p_{\rm id}(T,\mu^*=0)=0.975$
at $T=120$~MeV  and $\mu=0$. This behavior is similar to the pressure suppression due to the excluded volume effects with the hard-core radius of pion equal to $r\simeq0.3$~fm \cite{Poberezhnyuk2017}.

In the further discussion a temperature $T$ is the only independent thermodynamic variable, and $A$ is the one  free model parameter.
The system EoS depends then on a strength of attractive interactions described by numerical values of the parameter $A$.   
It is convenient to introduce a new dimensionless parameter $\kappa$ as
\eq{\label{kappa} 
\kappa ~ \equiv~ \frac{A}{A_{\rm cr}}~,
}
where $A_{\rm cr}$ is defined in Eq.~(\ref{Acr}).
For our choice of parameters the following 5 qualitatively different intervals of $\kappa$ are found:
\eq{\label{kappa-intervals}
[0,~\kappa_1],~~~ (\kappa_1,~ 1),~~~ (1,~\kappa_2),~~~ (\kappa_2,~\kappa_3),~~~(\kappa_3,~+\infty)~,
}
where
\eq{\label{kappa-values}
\kappa_1~\cong ~0.998,~~~~ \kappa_2~\cong~1.00017,~~~~ \kappa_3~=\frac{2}{\sqrt{3}}\cong 1.155~.
}

\vspace{0.3cm}
$ \bm{ 0\le \kappa \le \kappa_1: }$ {\bf No phase transitions. }

\begin{figure*}[t!]
\includegraphics[width=0.49\textwidth]{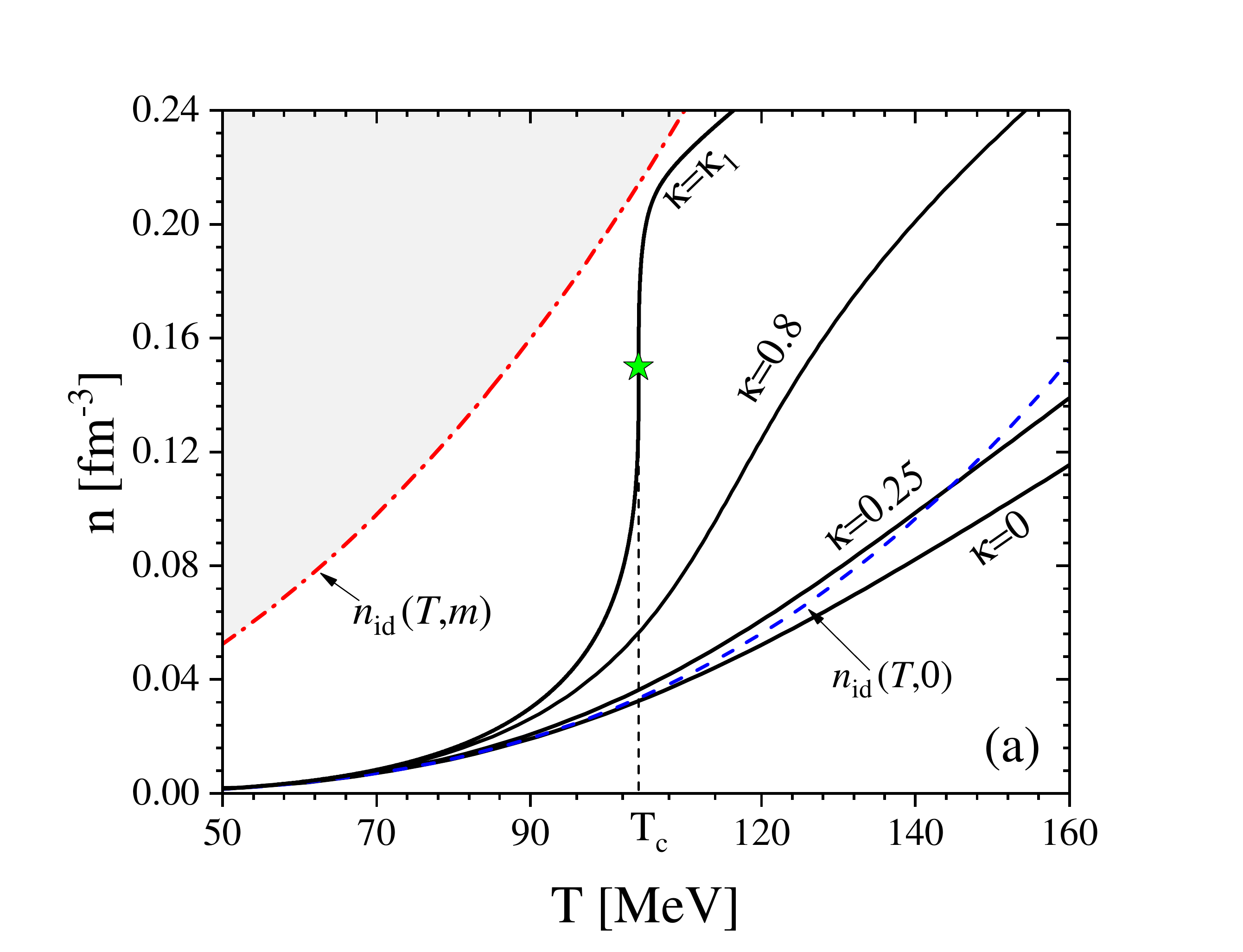}
\includegraphics[width=0.49\textwidth]{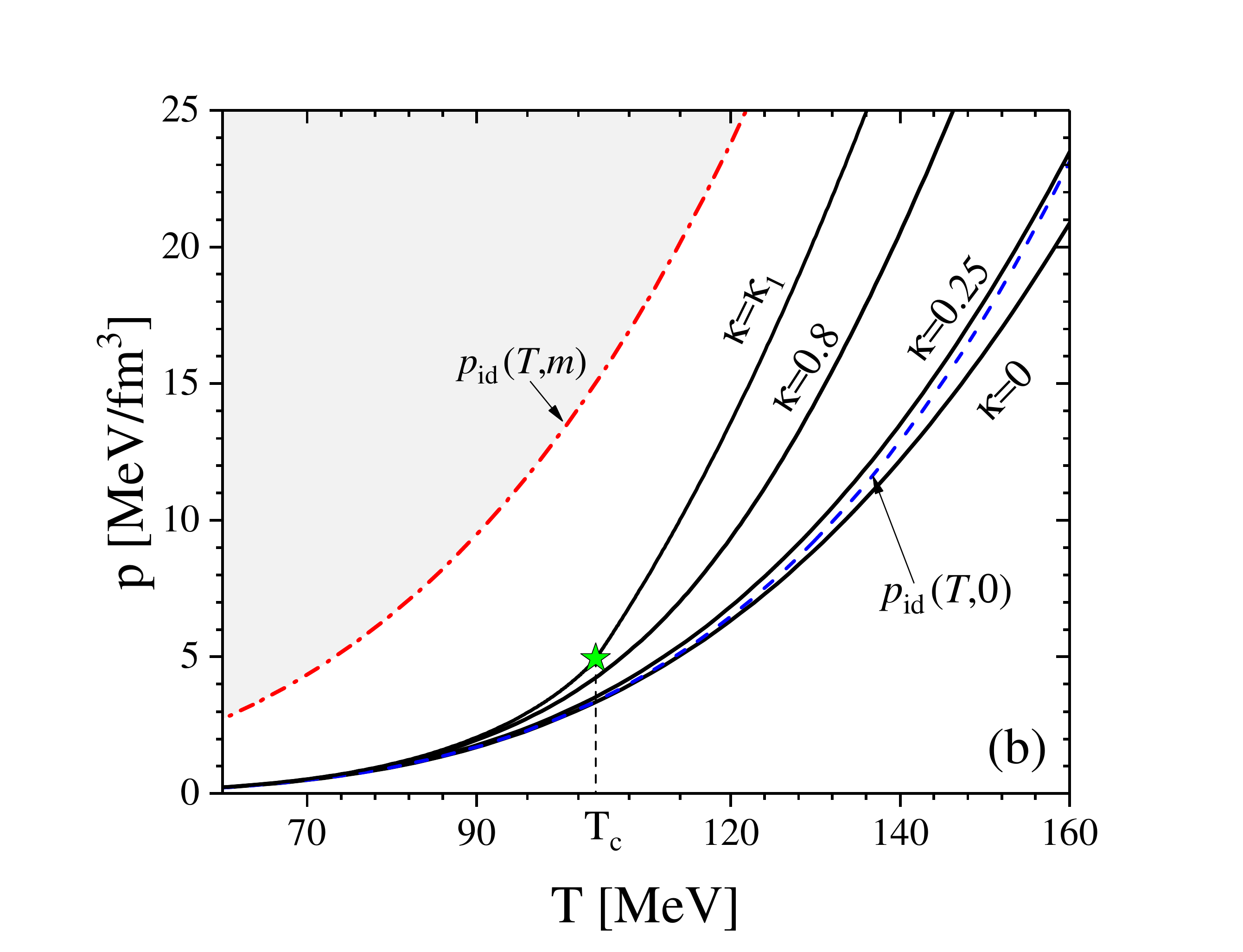}
\caption{\label{fig:Case1}The particle number density $n$ (a) and pressure $p$ (b) versus a temperature are shown by solid lines at different $\kappa \le \kappa_1$. Red dashed-dotted lines and blue dashed lines correspond to the ideal gas expressions (\ref{nid}) and (\ref{p-id}) with, respectively, $\mu^*=m$ and $\mu=0$. A star symbol denotes an inflection point of the $n=n(T)$ function on the line with $\kappa=\kappa_1$ at $T=T_c$. 
}
\end{figure*}

For values of $\kappa \in[0,\kappa_1]$ no BEC and/or phase transitions are possible. 
Functions $n=n(T)$ and $p=p(T)$ are shown for several fixed values of $\kappa \le  \kappa_1$ in Figs.~\ref{fig:Case1} (a) and (b), respectively. For a comparison, the ideal gas behavior with  
$\mu^*=0$, i.e., both $A=0$ and $B=0$, and $\mu^*=m$ are
also presented by dashed-dotted lines and dashed lines, respectively. 

\begin{figure*}[t!]
\includegraphics[width=0.49\textwidth]{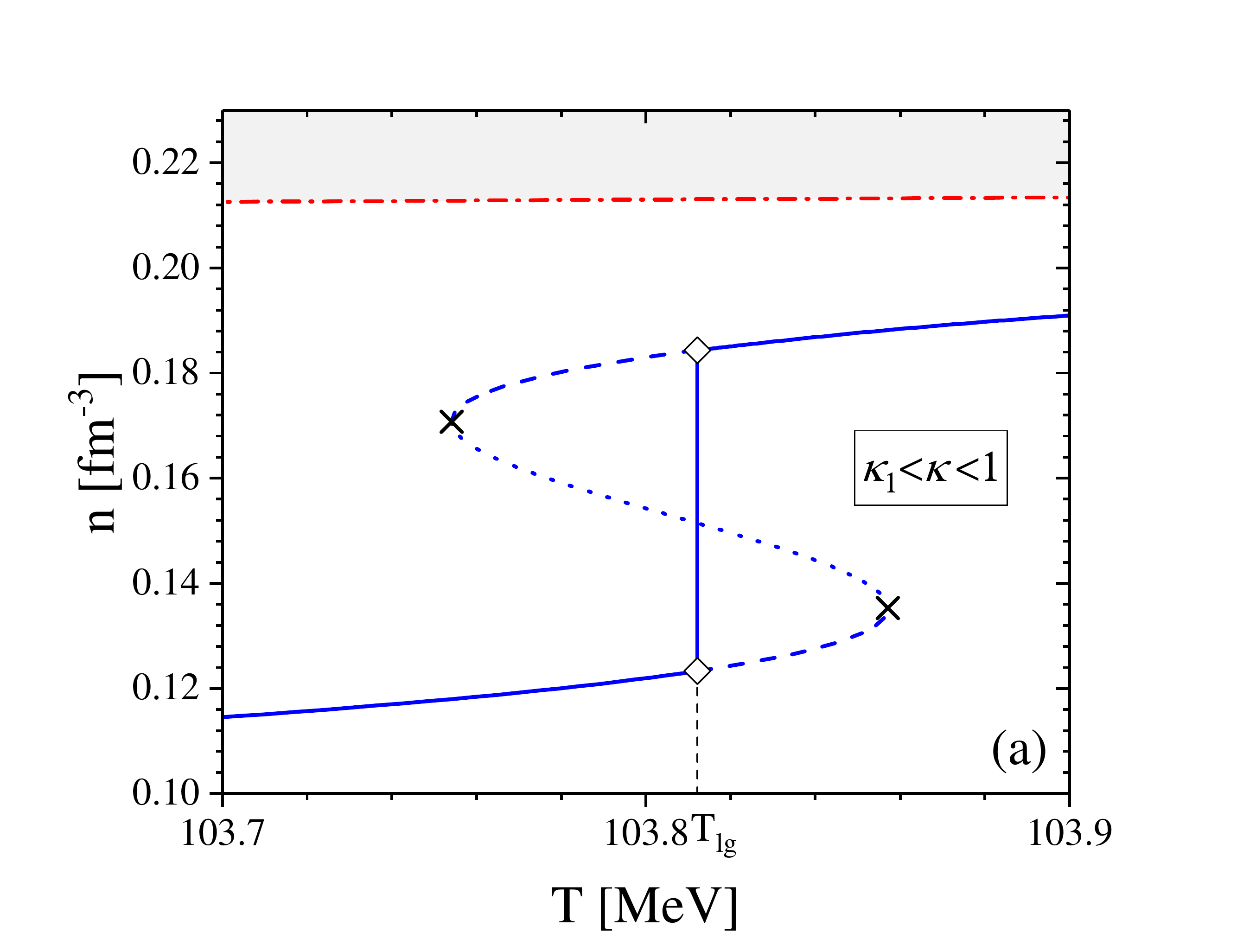}
 \includegraphics[width=0.49\textwidth]{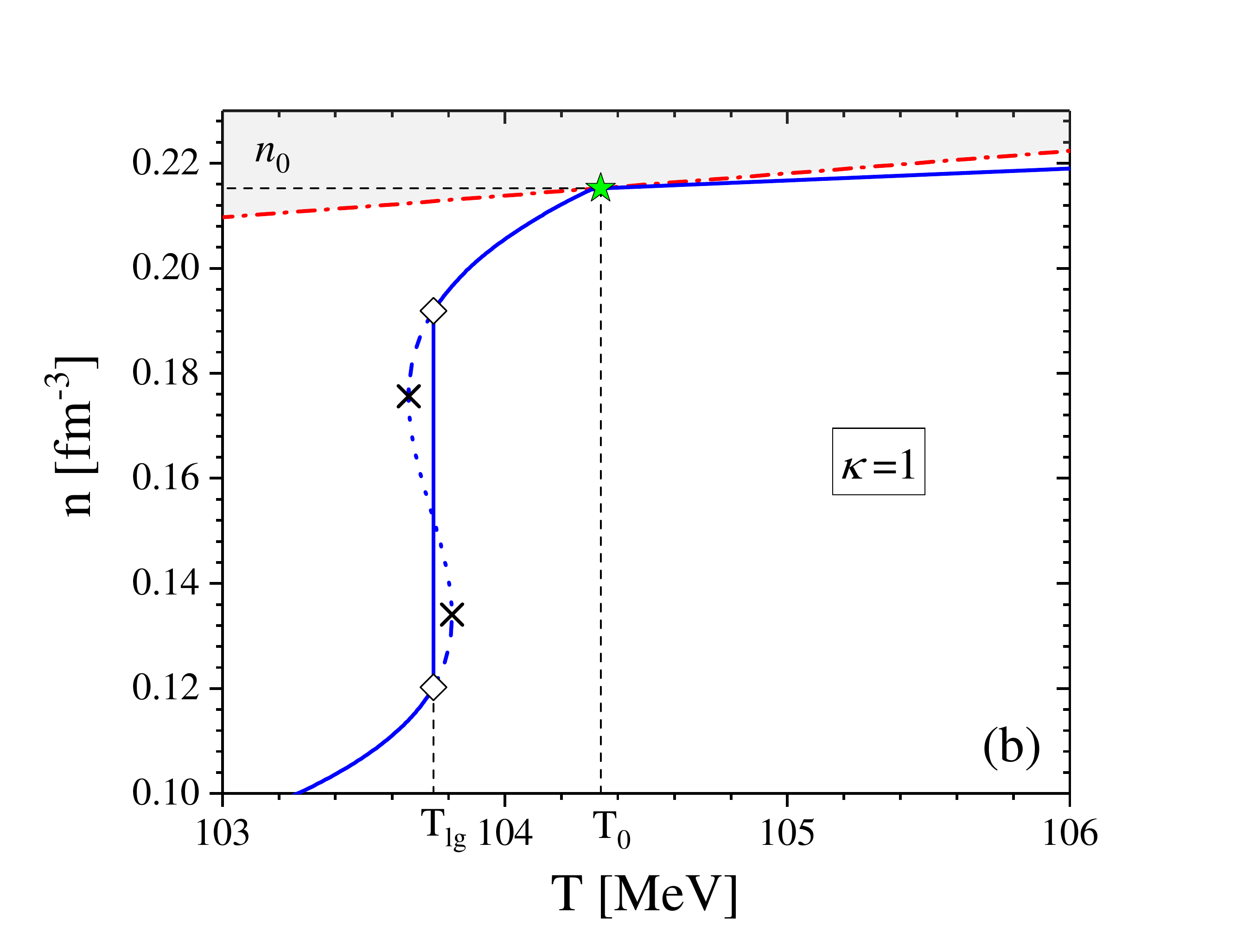}
\caption{\label{fig:pt1} 
The particle number density $n$ versus temperature for $\kappa_1< \kappa <1$ (a) and $\kappa=1$ (b) are shown by solid lines.  Dashed and dotted lines correspond to metastable  and unstable states of the system, respectively.
Vertical solid lines correspond to the liquid-gas mixed states of the LGPT at the phase transition temperature $T_{\rm lg}$. At $\kappa=1$ in (b) a star denotes an onset of the BEC at $n=n_0$
}
\end{figure*}

At small densities and large enough $\kappa$ the attractive effects dominate. This leads to $n(T,\mu^*)> 
n_{\rm id}(T,\mu^*=0)$ and $p(T,\mu^*)>p_{\rm id}(T,\mu^*=0)$. However, at very large $n$ a repulsive part of the $U(n)$ being proportional to $n^2$ always becomes larger than an attractive one proportional to $n$.

At $\kappa=\kappa_1$ the function $n=n(T)$ has an inflection point shown in Fig.~\ref{fig:Case1} by the star at $T=T_c$. At this point, $dn/dT=\infty$ that causes an infinite fluctuations of the number of particles as will be discussed in the next section.

\vspace{0.3cm}
$ \bm{\kappa_1< \kappa < 1:}$ {\bf Liquid-gas phase transition.}

For $\kappa \in(\kappa_1,1)$ the condition (\ref{Acr}) is not satisfied, thus, the BEC can not 
be realized. However, there is a possibility for the first order phase transition in this region of attractive forces.  
In some temperature interval $(T_1,T_2)$ the function $n=n(T)$ has three different solutions as it is shown in Fig.~\ref{fig:pt1} (a). 
A part of the $n(T)$ line with $dn/dT<0$ corresponds to the unstable solution, and it is shown in Fig.~\ref{fig:pt1} by the dotted line.  
Two other branches of $n=n(T)$ shown by dashed lines correspond to metastable states of the considered system.  
Statistical mechanics admits a possibility for an equilibrium between several different phases, e.g., two phases -- liquid and gas --  with different particle densities, $n=n_g$ and $n=n_l>n_g$. 
These phases can exist in the thermodynamic  equilibrium 
provided they have equal pressures (mechanical equilibrium), equal temperatures (thermal equilibrium), and equal  chemical potentials (chemical equilibrium).
For equal values of $T$ and $\mu=0$ in both phases, the phase with larger pressure is realized. And the phase transition temperature $T=T_{\rm pt}$ corresponds to the point where pressures of both phases become equal to each other. This is known as the Gibbs criterion for the first order phase transition. 
The vertical line in Fig.~\ref{fig:pt1} (a) corresponds to the mixed phase of a `gas' with density $n=n_g$ and `liquid' with $n=n_l$.  We use the standard names of `gas' and `liquid' for these phases with $n_g<n_l$ as it is usually done for the molecular gaseous and liquid systems. This phenomenon in our systems of interacting bosons  will also be denoted as the liquid-gas phase transition (LGPT).

Note that the states shown by the dashed lines in Fig.~\ref{fig:pt1}  can be realised in physical processes as metastable states. On the other hand, the states shown in Fig.~\ref{fig:pt1} by the dotted lines are unstable and they are fully forbidden.
In the considered case of the liquid-gas phase transition a Bose nature of constituents plays no essential role. Particularly, a conditions of the BEC with $\mu^*=m$ cannot be reached at any $T$. 
One finds that a picture of the considered here phase transition remains valid in the Boltzmann approximation, i.e, at $\eta=0$ in Eq.~(\ref{fk}).

\vspace{0.3cm}
$\bm{\kappa=1:}$ {\bf LGPT and an onset of the BEC.}

For $\kappa=1$, i.e., $A=A_{\rm cr}$, one finds $n_1=n_2=n_0$ from Eq.~(\ref{n12}). After the LGPT 
at $T=T_{\rm lg}$
the system reaches an onset of the BEC at temperature $T=T_0 > T_{\rm lg}$ and density $n=n_0$. This is just a point where the condition $\mu^*=m$ is satisfied, and an onset of the BEC is thus reached.
This is shown in Fig.~\ref{fig:pt1} (b).  The BC with $n_{\rm bc}>0$ cannot be formed: at both $T<T_0$ and $T>T_0$
one finds $\mu^* <m$. This case is especially interesting. The point $T=T_0$ and $n=n_0$ resembles a properties of a critical point.
Similar to the inflection point $T=T_c$ shown in Fig.~\ref{fig:Case1}  this is the second  point in our Bose system for which one observes an infinite particle number fluctuations (see
the next section).

\vspace{0.3cm}
$ \bm{1< \kappa< \kappa_2:}$ {\bf Two successive phase transitions.}

\begin{figure*}[t!]
\includegraphics[width=0.495\textwidth]{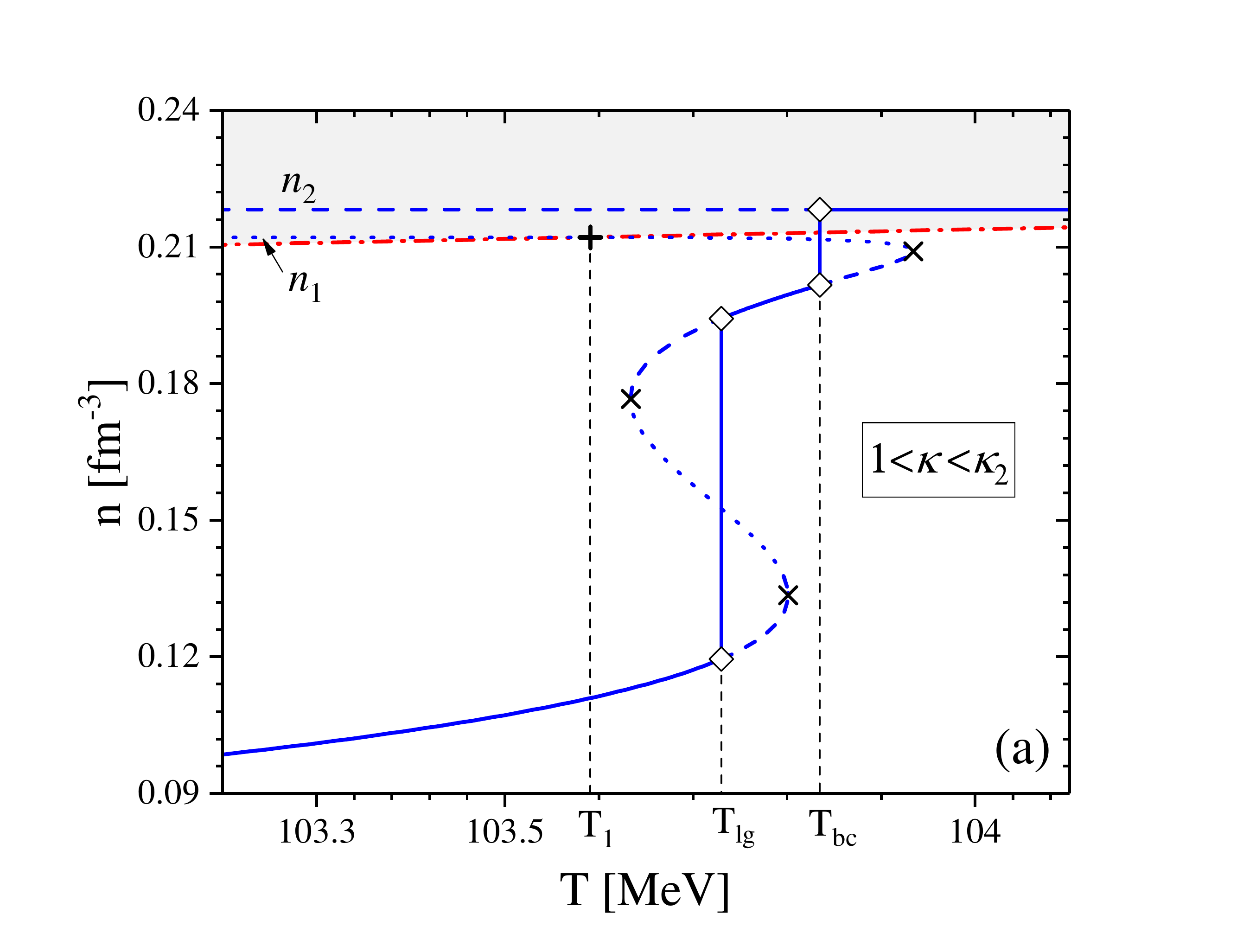}
\includegraphics[width=0.495\textwidth]{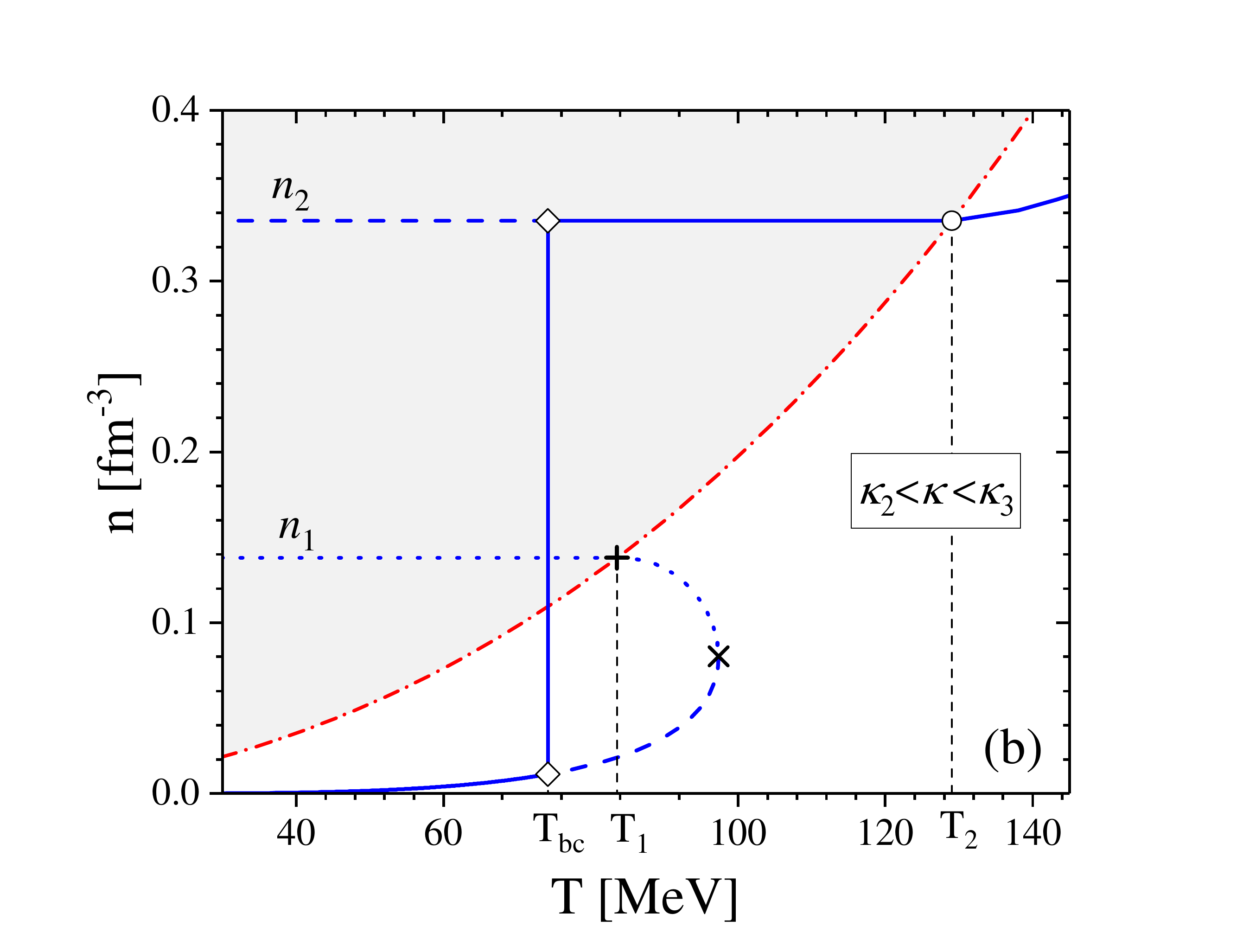}
\caption{\label{fig:pt2}
The particle number density $n$ versus temperature for $1<\kappa<\kappa_2$ (a) and $\kappa_2 <\kappa < \kappa_3$ (b) are shown by solid lines. Dashed and dotted lines correspond to metastable and  unstable states, respectively.
Vertical solid lines represent the mixed phases.
Two successive first order phase transitions are seen in 
(a): LGPT at $T=T_{\rm lg}$ and LGPT-BC at $T=T_{\rm bc}$. 
}
\end{figure*}

In this region of the attractive parameter the system EoS has two phase transitions.
This is illustrated in Fig.~\ref{fig:pt2}.
The first phase transition occurs at the temperature $T=T_{\rm lg}$, and it is similar to that of the LGPT shown in Figs.~\ref{fig:pt1} (a) and (b).
However, at $\kappa >1$ 
additional pure phase solutions with $n(T)=n_1$ and $n(T)=n_2$ exist. They  correspond to $\mu^*=m$  
and $n_{\rm bc}\ge 0$. 
One finds that the solution $n(T)=n_1$ is always an unstable one. The solution $n(T)=n_2$ is of a different nature. It is a metastable at small $T$, and becomes a stable one at $T= T_{\rm bc}$.
Both phase transitions shown in Fig.~\ref{fig:pt2} are the first order transitions and correspond to a jump in the particle number densities. A specific feature of the phase transition at $T=T_{\rm bc}$ is that the dense (`liquid') phase with $n_l=n_2$ includes the BC with $n_{\rm bc}>0$, whereas the diluted (`gas') phase with $n_g<n_1$ does not reach an onset of the BEC.

The second transition at $T_{\rm bc}$ is possible only due to the Bose statistics effects.
In a case of the Boltzmann approximation this type of transition is absent. 
For brevity we will call this type of a liquid-gas first order phase transition as the LGPT-BC to underline a presence of the BC  $n_{\rm bc}>0$  in a liquid phase at the transition temperature $T=T_{\rm bc}$.

\vspace{0.3cm}
$ \bm {\kappa_2<\kappa<\kappa_3:}$ {\bf LGPT-BC}.

At  $\kappa \in(\kappa_2,\kappa_3)$ the system evolves along the gas branch $n=n_{\rm id}(T,\mu^*)$
with $\mu^* <m$ up to the point $T_{\rm bc}$ where 
the liquid-gas phase transition with the BC $n_{\rm bc}>0$ in the liquid phase with $n_l=n_2$ takes place as shown in Fig.~\ref{fig:pt2} (b). This special possibility was previously considered 
in Refs.~\cite{Anchishkin_2019,Mishustin_2019}.

\vspace{0.3cm}
${ \bm{\kappa\ge \kappa_3=2/\sqrt{3}}:}$ {\bf Unstable vacuum.}

At $\kappa\rightarrow \kappa_3=2/\sqrt{3}$ from below, the temperature $T_{\rm bc}$ of the LGPT-BC moves to zero, and $T_{\rm bc}=0$ at  $\kappa = \kappa_3$. At $\kappa > \kappa_3$ the vacuum state $n=0$ at $T=0$ becomes metastable and transition to the stable state with $n_{\rm bc}=n_2$ at $T=0$ takes place. In this stable state, the pressure is positive, $p=p(T=0,n=n_2)=Bn_2^2n_0(\kappa-2\sqrt{\kappa^2-1})/3\geq 0$, and the energy density is negative, $\varepsilon =
\varepsilon(T=0,n=n_2)= - p(T=0,n=n_2) <0$. The exotic properties of this stable state with  $p=-\varepsilon$ at $T=0$ resemble those postulated for an EoS of the \textit{dark energy} in  models of the evolution of the Universe. 

\section{Particle number Fluctuations}\label{sec:fluc}
 The particle number fluctuations in our system can be characterized by susceptibilities
\eq{ \label{kn}
k_j~=~\left[\frac{\partial^j(p/T^4)}{\partial(\mu/T)^j}\right]_{\mu=0},~j=1,2,\ldots\,.
}
The scaled variance $\omega$ of the particle number distribution can be straightforwardly calculated as \cite{Satarov:2017jtu}
\be \label{omega}
\omega~=~\dfrac{k_2}{k_1}=~\omega_{\rm id}(T,\mu^*)~ \left[1~+~\omega_{\rm id}(T,\mu^*)\,\dfrac{n}{T}\dfrac{dU}{dn}\right]^{-1}~,
\ee
where $\omega_{\rm id}$ is the following ideal gas expression
\be \label{omega-id}
\omega_{\rm id}(T,\mu^*)~=~1~+~\frac{g}{2\pi^2n}\int\limits_{0}^{\infty}dk~k^2~f^2_{\rm k}(T,\mu^*)~.
\ee
A numerical value of $\omega=1$ corresponds to the Poisson particle number distribution. This result comes from Eq.~(\ref{omega}) for a classical gas of non-interacting  particles, i.e., $U(n)=0$. The Bose statistics and attractive interaction lead to an enhancement of $\omega$ and repulsive interaction to its suppression.   
\begin{figure*}[t!]
\includegraphics[width=0.495\textwidth]{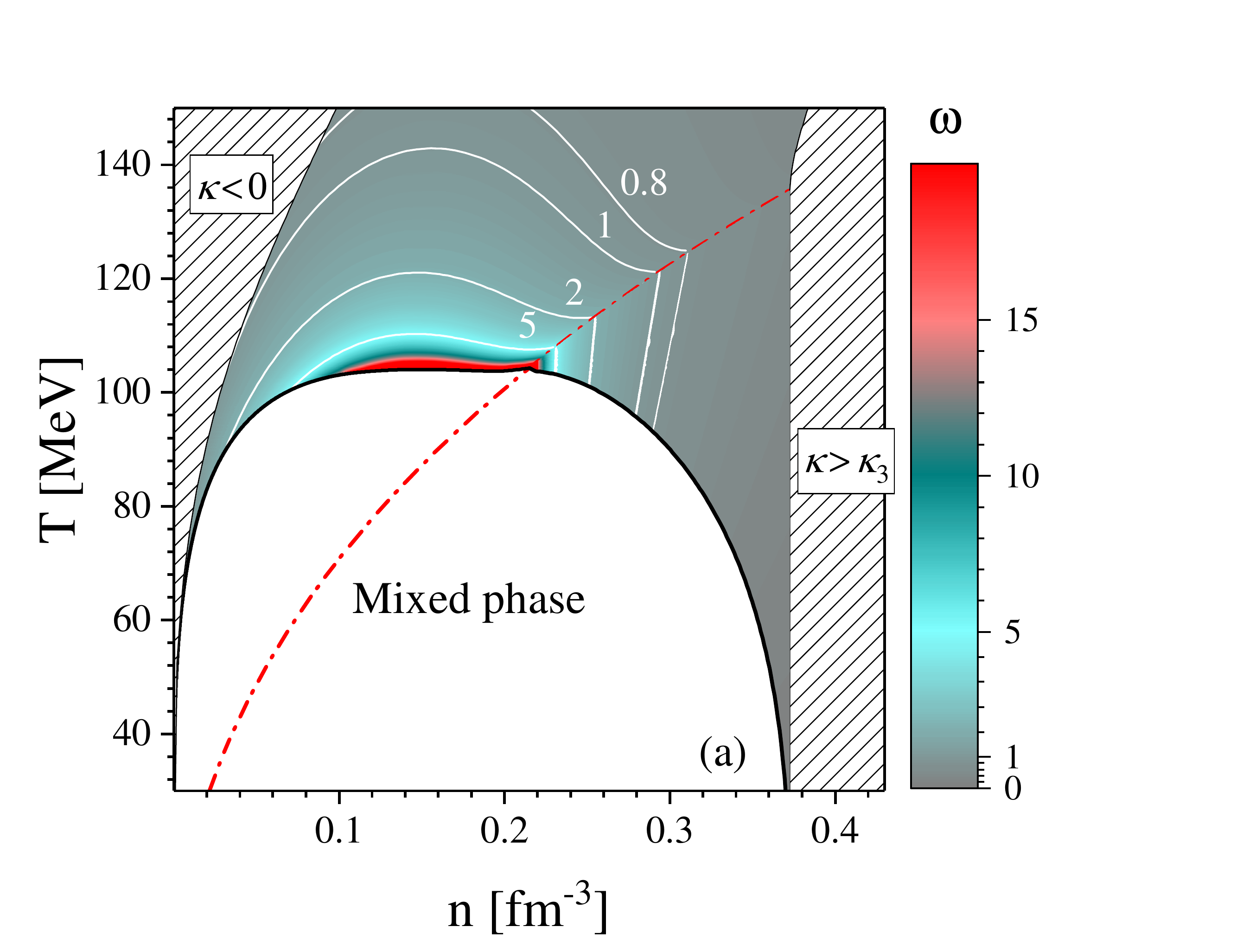}
\includegraphics[width=0.495\textwidth]{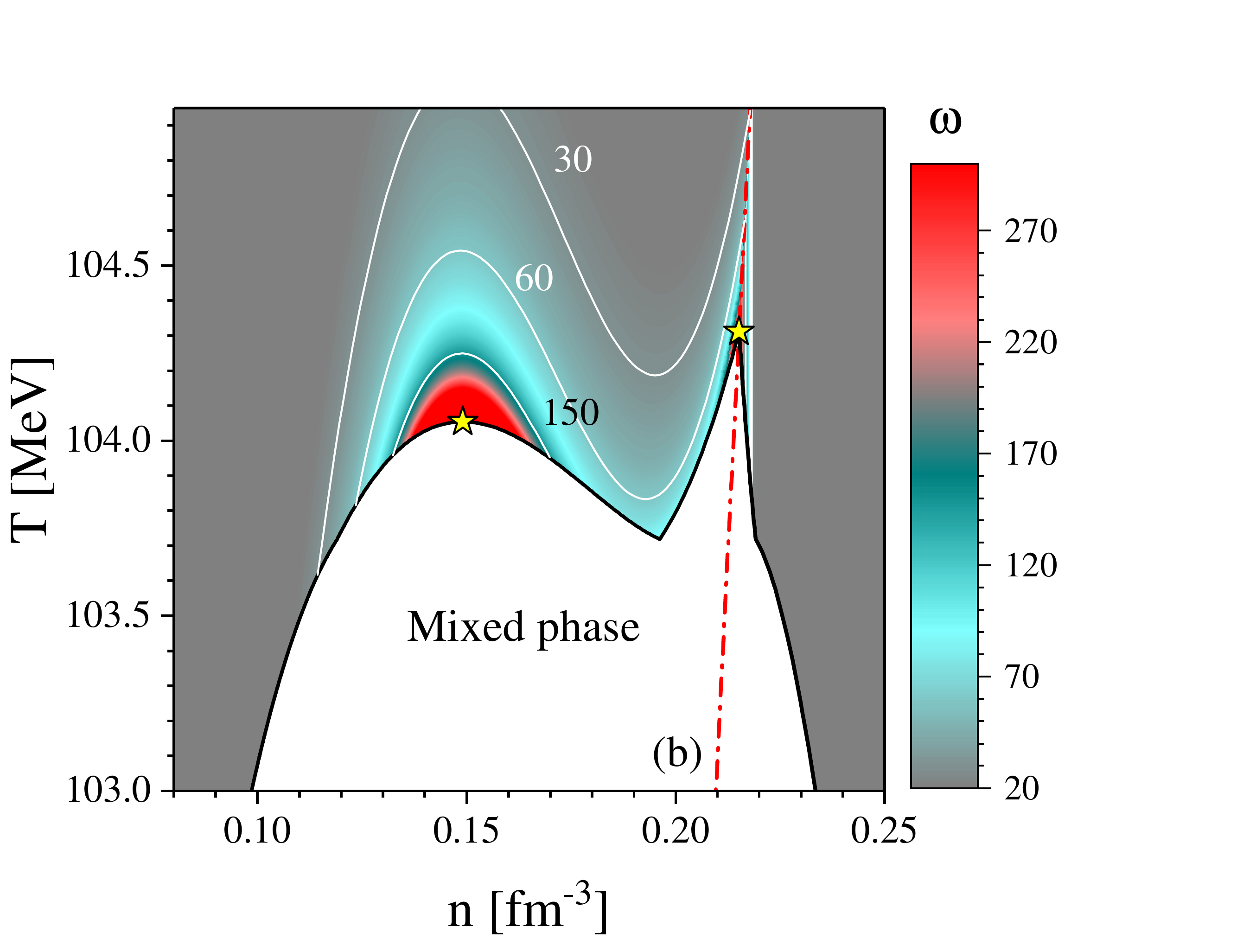}
\caption{
The scaled variance $\omega$ calculated in the pure gas or liquid phases is shown in the $(n,T)$-plane. This region correspond to the stable phase solutions $n=n(T)$.  The mixed phase is separated by the binodal curve shown by the black solid line. White lines denote the levels of constant values of $\omega$. Figure (b) tunes the scales to present  details of the critical point at $T=T_c$
for $\kappa=\kappa_1$ and an onset of the BEC at $T=T_0$ for $\kappa=1$. Both of these special points are denoted by the stars symbols.
}
\label{fig:omega}
\end{figure*}

Behavior of the $\omega$  calculated by Eqs.~(\ref{omega}) and (\ref{omega-id}) in the stable phases of the considered system is presented in the $(n,T)$-plane for different values of $\kappa$ in Fig.~\ref{fig:omega}. To guarantee  $\omega> 0$ as follows from its definition~(\ref{omega}), the following inequality should be satisfied
\eq{\label{omega-pos}
1~+~\omega_{\rm id}(T,\mu^*)\,\frac{n}{T}\frac{dU}{dn}~\ge 0~.
}
If the left hand side of Eq.~(\ref{omega-pos}) becomes equal to zero at some value of temperature, it leads to a divergence of the $\omega$ (\ref{omega}). This happens for $\kappa=\kappa_1$ at $T=T_c$ that is an inflection point of $n=n(T)$ function shown in Fig~\ref{fig:Case1} (a).

Another potential source of a divergence of the $\omega$ is the BEC. One observes that the scaled variance $\omega_{\rm id}$ become divergent  at $\mu^*=m$. This divergence is due to low momentum contribution of $f_k^2$ to the $k$-integral in Eq.~(\ref{omega-id}).
In this case one finds 
\be
\omega~=~\dfrac{T}{n}\left(\dfrac{dU}{dn}\right)^{-1}~.
\ee
A condition $\mu^*=m$ can be fulfilled either at $n=n_1$ or at $n=n_2$. A derivative $dU/dn$ is positive at $n=n_2$ and negative at $n=n_1$. Thus, the states with $n=n_2$ can be considered as the physical state, either stable or metastable. On the other hand, the states with $n=n_1$ lead to unphysical values of $\omega <0$. 

One special possibility at $\kappa=1$ corresponds to the point $T=T_0$ and $n\equiv n_0=n_1=n_2$
shown in Fig.~\ref{fig:pt1} (b).
At $T\rightarrow T_0$ one finds  $n \rightarrow n_0$ and  $\omega_{\rm id}(T,\mu^*)\to \infty$.  Taking into account the expansion of $n_{\rm id}(T,\mu^*)$ for $\mu^* \to m-0$~\cite{Begun:2008hq} one finds a leading term for $\omega_{\rm id}$ as
\eq{& \omega_{\rm id}(T,\mu^*) \stackrel{\mu^* \to m}{\simeq} \frac{g\,m^{3/2}T_0^2}{2\sqrt{2}\pi\,n(m-\mu^*)^{1/2}} \nonumber \\
&=~ \frac{g\,m^{3/2}T_{0}^{2}}{2\sqrt{2B}\pi n|n-n_0|}~,
}
where Eq.(\ref{MF-mu}) with $\mu=0$ was used at the last step. For the inequality~(\ref{omega-pos}) this gives 
\eq{
& 1+\omega_{\rm id}(T,\mu^*)\,\frac{n}{T}\frac{dU}{dn} \stackrel{\mu^* \to m}{\simeq}
1+\frac{g\,\sqrt{B}m^{3/2}T_0}{\sqrt{2}\pi}\, \frac{n-n_0}{|n-n_0|}\nonumber\\&\simeq 1+0.84 ~\text{sgn}(n-n_0) ~> ~ 0.
}
One  finally obtains 
\be
\label{cp}
\lim_{T \rightarrow T_0\pm 0} \omega~=+\infty~.
\ee
The both points, $T=T_c$ for $\kappa=\kappa_1$ and $T=T_0$ for $\kappa=1$, that correspond to $\omega=\infty$ are noted by the star symbols in Fig.~\ref{fig:omega} (b). 

\section{Summary}\label{sec:sum}
 A system of interacting bosons at finite temperatures and zero chemical potential was studied in the present paper within the Skyrme-like mean-field model. An interplay between attractive and repulsive interactions characterized by the model parameters $A$ and $B$ opens  possibilities for different types of the first order liquid-gas phase transition.
As a particular example an equilibrium system of  pions has been discussed. 
 At  different strengths of attractive forces characterized by dimensionless parameter $\kappa =A/(2\sqrt{mB})$ we found an abundance of the thermodynamic behavior: 1) critical point at $T=T_c$ for $\kappa=\kappa_1 \cong 0.998$;  2) liquid-gas phase transition for $\kappa_1 < \kappa <1$; (both cases 1 and 2 look similar to those in molecular system), 3) an onset of the Bose-Einstein condensation at $T=T_0$ for $\kappa=1$; 4) two successive phase transition at $1<\kappa < \kappa_2\cong 1.00017$, 5) a first order phase transition with the Bose condensate $n_{\rm bc}>0$ in the liquid phase for  $\kappa_2<\kappa< \kappa_3=2/\sqrt{3}\cong 1.155$  (in case 4, $n_{\rm bc}>0$ also appears in the liquid part during the second phase transition); 6) unstable vacuum for $\kappa >\kappa_3$.  
 Some of these situations, namely cases 1 and 2,  exist also in the Boltzmann approximation, i.e., they are  not sensitive to the effects of the Bose statistics. % 
 Effects of the Bose statistics introduce the new interesting possibilities, namely cases 3-6,
 when  non-zero value of the Bose condensate density exists in a liquid component of the mixed phase. 

The particle number fluctuation demonstrate an interesting behavior. Two special points with a divergence of the scaled variance $\omega$ were found: $T=T_c$ for $\kappa=\kappa_1$ and $T=T_0$ for $\kappa=1$. The first of these points resembles the critical point of the molecular systems, whereas the second one is a consequences of infinite fluctuations in the ideal Bose gas at the onset of the Bose-Einstein condensation with $\mu^*=m$ and $n_{\rm bc}=0$. Note that in the states with $n_{\rm bc}>0$ the anomalous fluctuations with $\omega=\infty$ are absent.  They are suppressed by the particle interactions.

Physical systems with $\mu=0$ resemble a  photon gas, when all intensive thermodynamic functions are defined by the system temperature only. As an example of such system one can consider neutral mesons. In addition to the Bose statistics  effects, these particle possess strong interactions that include both repulsive and attractive effects. A wide variety of qualitatively different scenarios can be then expected in such systems.

\vspace{0.3cm}
{\bf Acknowledgments.}
We are grateful to  I.N. Mishustin, L.M. Satarov, H. Stoecker, and V.I. Zhdanov  for fruitful discussions.
O.S.St. acknowledges the financial support from the scientific program “Astronomy and space physics” (Project N. BF19-023-01) of Taras Shevchenko National University of Kyiv.
The work of D.V.A. was supported by the National Academy of Sciences of Ukraine
by its priority project "Fundamental properties of the matter in the
relativistic collisions of nuclei and in the early Universe" (No. 0120U100935).
The work of M.I.G. was partially  supported
by the Program of Fundamental Research of the Department of
Physics and Astronomy of National Academy of Sciences of Ukraine.
\bibliography{references.bib}
\end{document}